\documentclass[conference]{IEEEtran}
\IEEEoverridecommandlockouts

\usepackage{cite}
\usepackage{amsmath,amssymb,amsfonts}
\usepackage{algorithmic}
\usepackage{graphicx}
\usepackage{textcomp}
\usepackage{xcolor}
\usepackage{soul}

\def\BibTeX{{\rm B\kern-.05em{\sc i\kern-.025em b}\kern-.08em
    T\kern-.1667em\lower.7ex\hbox{E}\kern-.125emX}}
\begin{document}

\title{An Appraisal-Based Approach to Human-Centred Explanations\\
}

\author{\IEEEauthorblockN{Rukshani Somarathna}
\IEEEauthorblockA{
\textit{Data61, CSIRO}\\
Canberra, Australia  \\
rsomarathna@acm.org}
\and
\IEEEauthorblockN{Madhawa Perera}
\IEEEauthorblockA{
\textit{Data61, CSIRO}\\
Canberra, Australia  \\
madhawa.perera@data61.csiro.au}
\and
\IEEEauthorblockN{Tom Gedeon}
\IEEEauthorblockA{
\textit{Curtin University}\\
Perth, Australia  \\
tom.gedeon@curtin.edu.au}
\and
\IEEEauthorblockN{Matt Adcock}
\IEEEauthorblockA{
\textit{Data61, CSIRO}\\
Canberra, Australia \\
matt.adcock@csiro.au}
}

\maketitle

\begin{abstract}
Explainability remains a critical challenge in artificial intelligence (AI) systems, particularly in high-stakes domains such as healthcare, finance, and decision support, where users must understand and trust automated reasoning. Traditional explainability methods—such as feature importance and post-hoc justifications—often fail to capture the cognitive processes that underlie human decision-making, leading to either too technical or insufficiently meaningful explanations. We propose a novel appraisal-based framework inspired by the Component Process Model (CPM) for explainability to address this gap. While CPM has traditionally been applied to emotion research, we use its appraisal component as a cognitive model for generating human-aligned explanations. By structuring explanations around key appraisal dimensions—such as relevance, implications, coping potential, and normative significance—our framework provides context-sensitive, cognitively meaningful justifications for AI decisions. This work introduces a new paradigm for generating intuitive, human-centred explanations in AI-driven systems by bridging cognitive science and explainable AI.
\end{abstract}


\section{Introduction}
With the rapid advancements in artificial intelligence (AI), there is a significant trend towards integrating AI into systems that interact with humans across domains such as healthcare, finance, and customer service, including social and collaborative robots. This integration has heightened the necessity for explainable AI (XAI)~\cite{cortinas2023toward, miller2019explanation}. A core challenge in these AI-driven systems is their tendency to operate as black boxes, withholding insights into how decisions are made and leaving users unaware of the reasoning behind specific outcomes. This opacity impedes trust, accountability, and effective decision-making, mainly when users must act upon system recommendations.

Existing XAI approaches, such as feature importance or model transparency methods, often focus on technical justifications, like identifying which features contributed to a decision. However, they tend to overlook human-like reasoning processes that align with human cognition-an important aspect when developing XAI systems that interact with users. To address this gap, we adopt the Component Process Model (CPM)~\cite{RN21Scherer, scherer2013coregrid}, a psychological framework that describes how humans evaluate events through five components: appraisal, motivation, physiology, expression, and feelings. Among these, cognitive appraisal is key for evaluating the relevance and meaning of events~\cite{RN21Scherer, RN269Scherer, RN34Reekum}. Thus, we propose an approach that emphasises a cognitively grounded form of explainability, mirroring human judgment by focusing on appraisals, such as relevance, implications, coping potential, and normative significance.

As conceptualised by Scherer’s CPM, appraisals form the foundation of emotional and cognitive evaluations. They describe how individuals assess the significance of an event, its potential consequences, their ability to cope with it, and how it aligns with personal or societal norms~\cite{RN21Scherer, scherer2013coregrid, RN269Scherer, RN34Reekum}. Motivated by the breadth of research that uses CPM, particularly its application in emotion recognition~\cite{RN34Reekum, somarathna2023emostim, RN590Menetrey}, this work expands the use of the appraisal component as a proxy model for explainability in decision-making contexts. Instead of providing generic explanations or post-hoc justifications, the system can dynamically maps its decisions to appraisal objectives such as relevance (\textit{whether an event is relevant}), implications (\textit{what are the consequences of the event?}), coping potential (\textit{how it can be coped with?}), and normative significance (\textit{does the event align with the person’s values and norms?}). This enables accurate and meaningful explanations to users, enhancing trust and engagement across various domains. For example, instead of simply stating that an event was classified as `anger,' the system can explain that the classification was based on the event being appraised as violating norms, having negative consequences, and being uncontrollable.

In summary, this paper contributes a conceptual and computational framework that demonstrates how appraisal-based reasoning can be operationalised as a data-driven, real-time approach to generate human-centred explanations in AI-driven systems. Grounded in the CoreGRID appraisal dimensions~\cite{scherer2013coregrid}, our framework illustrates how AI decisions can be explained in ways that align with human cognitive processes across different decision-making contexts. We demonstrate the feasibility of this framework through a case study in a simulated natural language processing (NLP) environment, illustrating how appraisal-based reasoning can enhance interpretability in AI decision-making. It can be applied in domains such as social robotics, affective virtual agents, autonomous vehicles, healthcare diagnostics, educational technologies, and intelligent decision support systems, where understanding user intent, emotional state, and contextual factors is critical for building trust and enhancing interaction quality.

\section{Background and Related Work}

\subsection{Background}

\subsubsection{Component Process Model (CPM)}
The CPM offers a comprehensive framework for understanding emotions as responses to stimuli, structured around five key components: \textit{appraisal, motivation, expression, physiology}, and \textit{feeling} \cite{RN21Scherer, RN590Menetrey, somarathna2023emostim, somarathna2022multiAJCAI}. Among these, the appraisal component is foundational, involving cognitive evaluations determining an event's significance to well-being. Specifically, appraisal assesses relevance (\textit{Is the event relevant?}), implications (\textit{What is its impact?}), coping potential (\textit{Can I overcome the consequences?}), and normative significance (\textit{Does it align with my values?}) \cite{zhan2023evaluating, RN269Scherer, RN34Reekum, somarathnaexploring}. These evaluations guide emotional responses, making the appraisal process central to emotional experience and subsequent processes \cite{RN269Scherer, RN34Reekum}. 

By focusing exclusively on the appraisal component, we isolate the core cognitive mechanisms behind emotional reactions, enabling structured and interpretable explanations in affective systems. This focus allows explanations to mirror human-like reasoning and improve their transparency. Moreover, the appraisal process forms the basis for computational assessment within CPM, making it valuable for generating explanations that align with human thought processes.

\subsubsection{CoreGRID Appraisal Questionnaire}  
The appraisal section of the CoreGRID captures cognitive evaluations associated with emotional responses, focusing on the appraisal dimensions of relevance, implications, coping potential, and normative significance \cite{scherer2013coregrid, somarathna2023emostim, RN590Menetrey, somarathna2022multiAJCAI, SomarathnaPerCom9767281}. For example, relevance refers to whether an event is important to goals, and implications address the event's negative consequences.
These features offer a rich dataset for understanding the cognitive mechanisms underlying emotional experiences. By aligning appraisal-focused questions with these features, the CoreGRID dataset enhances the interpretability of AI-generated explanations, bridging the gap between AI outputs and human-like reasoning in affective computing.

\subsection{Related Works}
Human-centred AI, using a human-in-the-loop approach, enhances system outputs by integrating user feedback, improving both performance and explainability \cite{giaccardi2020technology}. This paradigm, grounded in human factors and cognitive science, facilitates AI systems that align with user needs, particularly in sensitive contexts like process industries \cite{kotriwala2021xai, suffian2023investigating}.

In the domain of explainable affective computing, transparency and trust are paramount. Recent reviews highlight the significance of trust, faithfulness and causality in affective computing, while recognising that the path toward developing explainable and transparent systems has already been paved \cite{cortinas2023toward, johnson2024explainable}. These methods clarify how systems interpret complex, multimodal, and time-dependent data. However, challenges persist, including managing contextual influences, ensuring privacy, and aligning explanations with user needs. Additionally, making a model transparent does not guarantee it will be entirely understandable to users, highlighting the need for explanations that are not only accurate but also cognitively accessible \cite{suffian2023investigating}. While basic input-output explanations are possible, more refined approaches are necessary to create scalable, user-centred systems. 

Incorporating physiological signals, such as facial expressions and eye tracking, improves the explainability of affective systems by providing interpretable insights into human behaviour \cite{mouakher2024explainable, bekler2024assessing}. Techniques like multitask predictions and componential modelling enhance the reliability of facial emotion analyses, and eye-tracking data interpreted with XAI methods like LIME~\cite{ribeiro2016should} and SHAP~\cite{lundberg2017unified} can clarify system explanations \cite{guerdan2021toward, bekler2024assessing}. Also, \cite{liu2024explainable} employed gradient-based visualisation techniques to interpret eye-tracking data. Furthermore, PeCoX’s cognitive XAI framework \cite{neerincx2018using}, which incorporates goals, beliefs, and emotions to explain agent behavior, aligns with the BDI (Belief-Desire-Intention) models \cite{harbers2010design} by providing explanations from the intentional stance. Integrating emotional factors offers a more nuanced understanding of the agent's decision-making process. However, while this approach is effective within the scope of goals, beliefs, and emotions, a more comprehensive understanding of user needs, encompassing a broader range of dimensions, would offer even greater insights \cite{neerincx2018using}. Despite progress, a gap remains in generating human-centred, contextually relevant explanations.

\textbf{Predictability} is crucial for building trust in human-AI interactions, as users develop expectations based on past experiences \cite{schadenberg2021see}. Enhancing predictability in explanations helps users form accurate expectations and improves interaction quality. For instance, users may perceive an autonomous vehicle as more predictable if it mimics human-like behaviour, even if its actions are less predictable \cite{driggs2016communicating}.

\textbf{Causality} and \textbf{control} are also critical for adequate explanations, as they help users differentiate between correlations and the proper drivers behind system behaviour \cite{arnold2021explaining}. People naturally seek the causal history of events to build stable mental models for prediction and control \cite{miller2019explanation, hilton2007course}. For example, users may react differently to an AI assistant’s actions depending on whether the event was caused by their own input or an autonomous decision. The perception of \textbf{power status} and \textbf{controllability} plays a key role in designing XAI systems \cite{ha2022examining}. In contexts like traffic enforcement, a high power status may lead users to view the system's results as manipulable, undermining trust in its reliability. Conversely, in supportive contexts, such as recommendation systems, power status may have a less significant effect or even the opposite. Emphasising stability and controllability in XAI explanations can enhance trust and acceptance. Moreover, understanding the \textbf{purpose} and \textbf{prospective action} behind system behaviour helps align actions with broader user goals and anticipated outcomes, improving alignment with user expectations \cite{arnold2021explaining}.

Incorporating \textbf{normative significance} into XAI frameworks is crucial for explaining whether system behaviour aligns with personal ideals, socially accepted norms, or beliefs \cite{arnold2021explaining, sado2023explainable, harbers2010design}. This is particularly important in affective contexts, where users assess whether the system's actions align with their expectations. Previous work shows that people interpret intentional actions through reason-based explanations rooted in beliefs, desires, and values, which help judge the appropriateness of behaviour according to shared standards \cite{miller2019explanation}.

Traditional XAI methods, such as LIME~\cite{ribeiro2016should} and SHAP~\cite{lundberg2017unified}, and counterfactuals, often focus on feature importance or alternative scenarios, but they fail to account for the subjective, contextual, and emotional aspects of decision-making \cite{guerdan2021toward}. Key factors like personal experiences, cultural norms, and situational influences are primarily overlooked, creating a gap in generating human-aligned explanations.

The CPM provides a structured approach for understanding emotions and behaviours by examining the interplay between appraisal, motivation, physiology, expression, and feeling \cite{delplanque2009sequential}. Recent studies show the potential of using appraisal dimensions to guide explainability in affective systems, aligning system behaviours with users' cognitive and emotional processes \cite{zhan2023evaluating}. For example, a key limitation in human-robot collaboration is the lack of focus on helping humans understand robot actions and intentions, which can reduce interaction quality and performance \cite{kumar2024exploratory}. To address this, frameworks can define \textbf{action concepts}—specific robot behaviours for explanation—and \textbf{input concepts}, high-level attributes of inputs influencing those actions \cite{sagar2024trustworthy}. For example, if an autonomous vehicle fails to yield, understanding which attributes of the situation, such as the perceived urgency based on the speed or type of the oncoming vehicle, influenced that action can help clarify the system's reasoning. This kind of contextual evaluation improves transparency and understanding. The CPM could help bridge the gap by focusing on action and input concepts in human-robot interactions, improving transparency and understanding of system decisions. 

The 21 appraisal items from the CoreGRID provide a comprehensive framework for generating human-centred explanations by aligning system behaviours with key emotional constructs such as event predictability, self-causation, relevance to personal goals, emotional valence, dominance, coping potential, normative significance, and urgency \cite{scherer2013coregrid}. These constructs help map system actions to users' cognitive and affective processes, enhancing interpretability, trust, and engagement. In contrast, traditional methods like LIME~\cite{ribeiro2016should} and SHAP~\cite{lundberg2017unified}, and counterfactuals often overlook the nuanced interplay of cognitive-emotional processes. While feature attribution methods focus on quantifying feature importance, they fail to capture decision-making’s subjective, contextual, and emotional facets, highlighting the need for models that integrate appraisal dimensions for more human-aligned explanations \cite{guerdan2021toward, kaptein2017role}.

A key challenge in current XAI models is tailoring explanations to individual user needs, while considering contextual and cognitive factors \cite{suffian2023investigating}. Many existing systems struggle to adapt explanations to diverse cognitive and social contexts, resulting in a misalignment between the explanations provided and the user's specific goals. Moreover, current models often neglect valuable insights from cognitive science and psychology, essential for enhancing user comprehension. While interdisciplinary approaches, such as those proposed by \cite{miller2019explanation}, advocate for integrating social sciences into XAI, these factors remain underexplored in current models, limiting their real-world applicability. Ultimately, there is a pressing need for XAI systems that more effectively address user goals, cognitive processes, and social dynamics, thereby improving both interpretability and overall utility.

\section{Framework and Methodology}
We present a conceptual and computational framework that uses the appraisal component of the CPM to generate interpretable explanations for autonomous decisions. This approach treats appraisals as explanation primitives—cognitively grounded building blocks that allow autonomous systems to communicate their reasoning in terms familiar to human users. The framework is domain-agnostic and applicable wherever autonomous systems make decisions that affect humans, particularly in contexts where trust, interpretability, and user actionability are critical.

\begin{figure*}[ht!]
\centering
\includegraphics[width=0.95\linewidth]{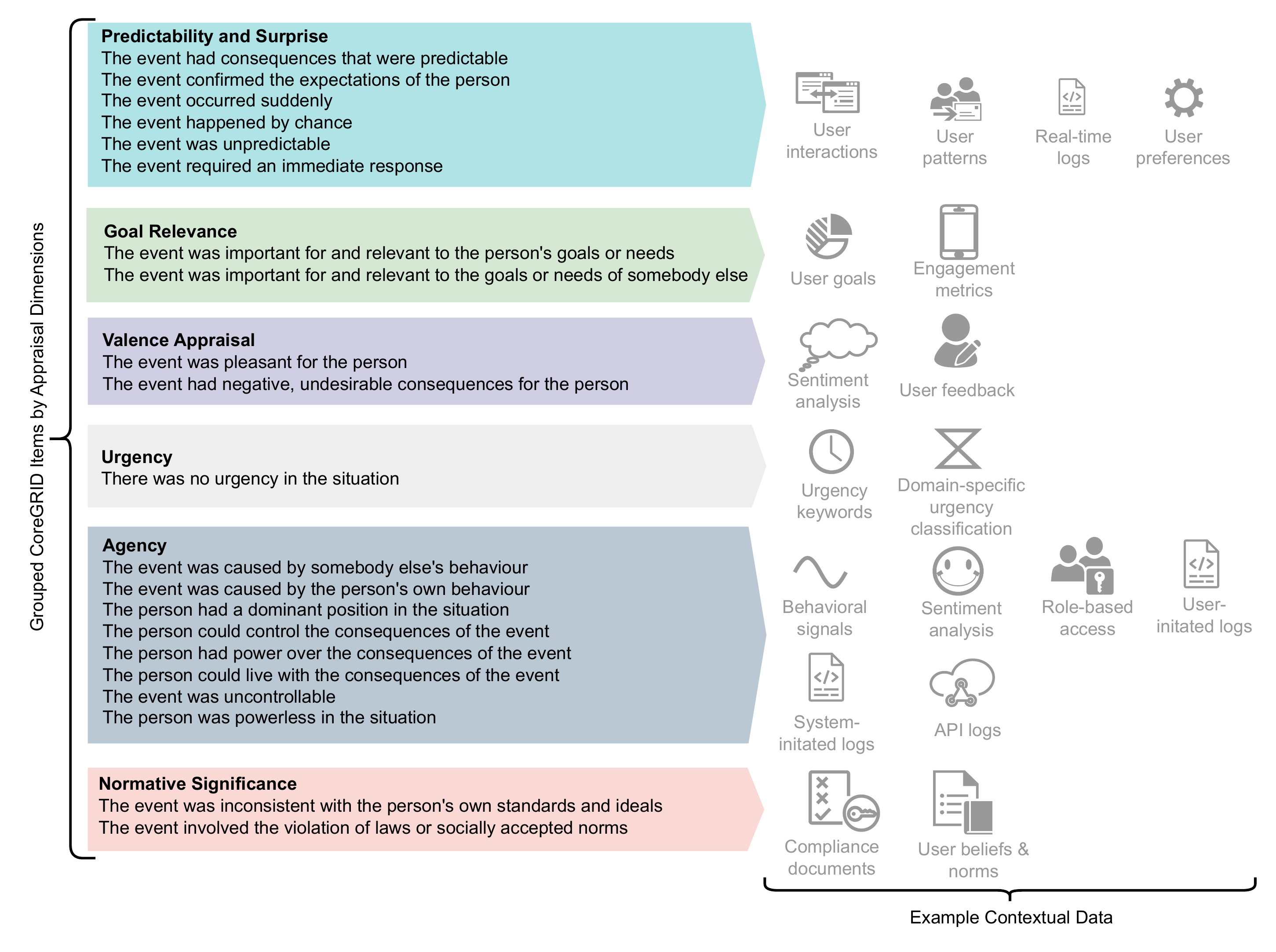}
\caption{An overview of our appraisal-based approach, where CoreGRID appraisal items are categorised into five appraisal dimensions, with contextual data provided for a deeper understanding of each dimension.}
\label{fig:Framework}
\end{figure*}

\subsection{Appraisal-Focused Explanation Generation}
As illustrated in Figure~\ref{fig:Framework}, this framework's core is a fundamental insight from the CPM: human evaluations of events are structured along key appraisal dimensions, such as relevance, causation, implications, and normative alignment. Conventional AI explainability approaches often rely on technical components, such as model weights, gradients, or feature importance, which remain inaccessible to non-expert users. Instead, we propose a shift towards human-interpretable appraisals, where explanations are framed in cognitive evaluations that naturally align with human reasoning. We used 21 appraisal items from CoreGRID~\cite{scherer2013coregrid} to structure our framework, ensuring a comprehensive assessment of explainability within our CPM-based approach. These items were grouped into core appraisal dimensions—such as \textbf{predictability and surprise, relevance, valence appraisal, urgency, agency}, and \textbf{normative significance}—to support systematic, cognitively grounded evaluations of system-generated explanations. Structuring AI explanations around appraisal dimensions ensures they are intuitive, meaningful, and aligned with user expectations, enhancing interpretability, trust, and usability in AI-driven systems.

\subsection{Key Appraisal Dimensions for Explainability}
\textbf{Predictability and Surprise} play a central role in how humans process events. When an event aligns with prior experiences, it reinforces trust in the system; additional clarification is needed when it deviates. AI-generated explanations should explicitly acknowledge whether an event was expected or unexpected. Following \cite{scherer2013coregrid}, we grouped six CoreGRID items to examine how predictability and surprise interact in shaping human experiences. These included two expected/familiar items—\textit{the event had predictable consequences} and \textit{the event confirmed the expectations of the person}—and three unexpected/chance items—\textit{the event occurred suddenly}, \textit{the event happened by chance}, and \textit{the event was unpredictable}, as well as \textit{the event required an immediate response}. This combined category provides a valuable framework for evaluating system behaviour and building a user knowledge base through data-driven appraisal. Techniques such as user interactions, preferences and pattern analysis, real-time event logs of event likelihood can be used to infer and update user expectations. Feeding this information into the system enables AI to explain outcomes more clearly, reduce uncertainty, and personalise interactions based on individual predictability thresholds.

\textbf{Goal Relevance} ensures that AI decisions are assessed based on their alignment with user needs. Humans interpret events through their significance to personal or external objectives, and AI explanations should mirror this interpretative process. In this context, the two CoreGRID items—\textit{the event was important for and relevant to the person's goals or needs} and \textit{the event was important for and relevant to the goals or needs of somebody else}—are particularly relevant. To operationalise this, we incorporate user feedback, engagement metrics, and user-specific logic, enabling the system to convey whether an outcome aligns with user-defined goals. By building a user dataset based on these factors, AI systems can better tailor their actions and explanations to support individual goals, thereby enhancing both decision-making and explainability in a user-aligned manner.

\textbf{Valence Appraisal} determines whether an event has positive or negative implications for the user. This dimension is crucial in shaping user trust and engagement, as individuals evaluate whether a system’s decision benefits or harms them. Using analyses such as sentiment analysis, feedback, and contextual cues, automatic systems should consider user-specific positive valence factors, such as preferences, goals, and emotional responses. This alignment helps ensure decisions resonate more strongly with users, improving AI behaviour's perceived soundness and explainability. To capture this, we incorporate the following CoreGRID items: \textit{the event was pleasant for the person} and \textit{the event had negative, undesirable consequences for the person}.

\textbf{Urgency} is crucial in time-sensitive decision-making, as some events require immediate action while others allow for flexibility. AI-generated explanations should convey the urgency of an event by integrating domain-specific urgency classifications, real-time monitoring of user engagement, and keyword-based urgency detection.  To support this, we use the CoreGRID item: \textit{there was no urgency in the situation}. Providing this information empowers users to make informed decisions under time constraints.

\textbf{Agency} and \textbf{Causation} play a crucial role in shaping user perceptions of control, responsibility, and trust in AI systems. To capture agency, we utilise CoreGRID items reflecting dominance, control, power, and coping potential—such as \textit{the person had a dominant position in the situation}, \textit{the person could control the consequences of the event}, and \textit{the person could live with the consequences of the event}—as well as low-agency indicators like \textit{the event was uncontrollable} and \textit{the person was powerless in the situation}. When users sense a lack of control, their trust in the system can be undermined; thus, AI explanations should convey the degree of user agency by incorporating behavioural cues, interaction history, and system affordances. In parallel, causation addresses whether user actions or autonomous system processes drove an outcome. Items like \textit{the person's behaviour caused the event} versus \textit{somebody else's behaviour caused the event} support this distinction. Using role-based access, event logs, API traces, and user and system interaction data, AI systems can offer clear causal attributions, helping users understand their influence over outcomes and reinforcing transparency and accountability.

\textbf{Normative Significance} ensures that AI decisions align with legal, ethical, and societal standards, guided by two CoreGRID items: \textit{whether an event is inconsistent with a person’s standards and ideals}, and \textit{whether it involves the violation of laws or socially accepted norms}. Users assess events based on their adherence to established norms, making it essential for AI systems to clarify whether an action complies with regulations and ethical principles. This can be achieved by referencing policy compliance documents, predefined ethical frameworks, and legal databases that map violations to industry standards. Providing norm-based justifications enhances transparency and fosters trust in AI decision-making.

\subsection{Case Study}

\begin{figure*}[htbp]
\includegraphics[width=1\linewidth]{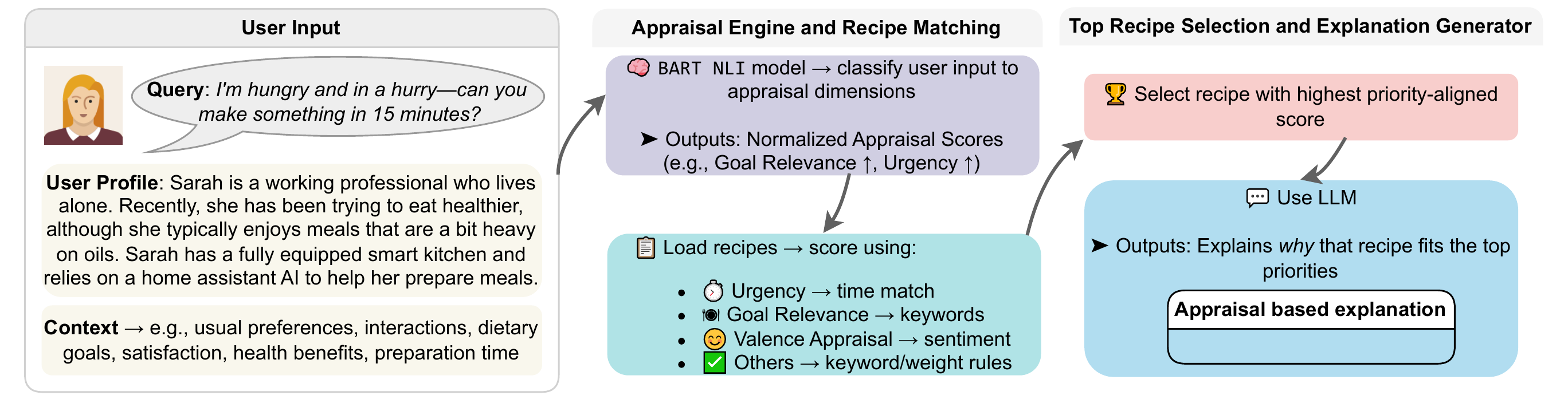}
\caption{Multi-stage explanation generation pipeline for applying the appraisal-based approach in meal recommendation.}
\label{fig:Usecase1}
\end{figure*}

To demonstrate how the appraisal component of the CPM can inform the generation of interpretable explanations in autonomous systems, we present a case study in the context of meal recommendation. This example illustrates how appraisal dimensions can be dynamically inferred from user context and translated into tailored, context-aware explanations that reflect the user’s affective and practical priorities. We developed a multi-stage explanation generation pipeline that integrates structured user modelling with large language models (LLMs) to produce human-centric, emotionally resonant explanations. The pipeline is grounded in six core appraisal dimensions: \textit{predictability and surprise, goal relevance, valence appraisal, urgency, agency}, and \textit{Normative Significance}. These dimensions are operationalised to interpret persistent user preferences and immediate situational cues.

The process begins by constructing a unified user context from a long-term (data-driven) profile and a natural language query. As given in Figure~\ref{fig:Usecase1}, the user, Sarah, is a busy, health-conscious individual seeking a quick and satisfying dinner. Her query, \textit{``I'm hungry and in a hurry, can you make something in 15 minutes?''} is interpreted in light of her broader lifestyle preferences. The system encodes this context into appraisal-relevant expectations using heuristics such as keyword detection, sentiment analysis, and time constraints (Refer to Supplementary Figure 5).

To assess which appraisal dimensions are most salient in the current context, we employ a natural language inference model (facebook/bart-large-mnli). The model compares the composite user input against statements representing each appraisal dimension and returns confidence scores. These are normalised and used to prioritise the appraisal dimensions that best capture the user’s momentary concerns—for instance, high urgency, goal relevance and predictability.

Subsequently, a curated set of candidate recipes (generated via GPT-4o) is evaluated based on how well each aligns with the prioritised appraisals. For example, we compute appraisal-aligned attribute scores for each recipe by analysing textual descriptions and metadata (e.g., cooking duration, ingredients, tone). Custom scoring functions are used per dimension: Urgency is determined through estimated preparation time and urgency-related keywords, while Valence Appraisal draws on sentiment cues. Recipes are then ranked according to their composite appraisal alignment, ensuring the final selection is practically appropriate and emotionally attuned to the user’s needs. (Simplified scoring methods are used for this paper, as the development of detailed scoring is outside the scope of our work; we focus on presenting the framework.)

\begin{figure}[htbp]
\includegraphics[width=1\linewidth]{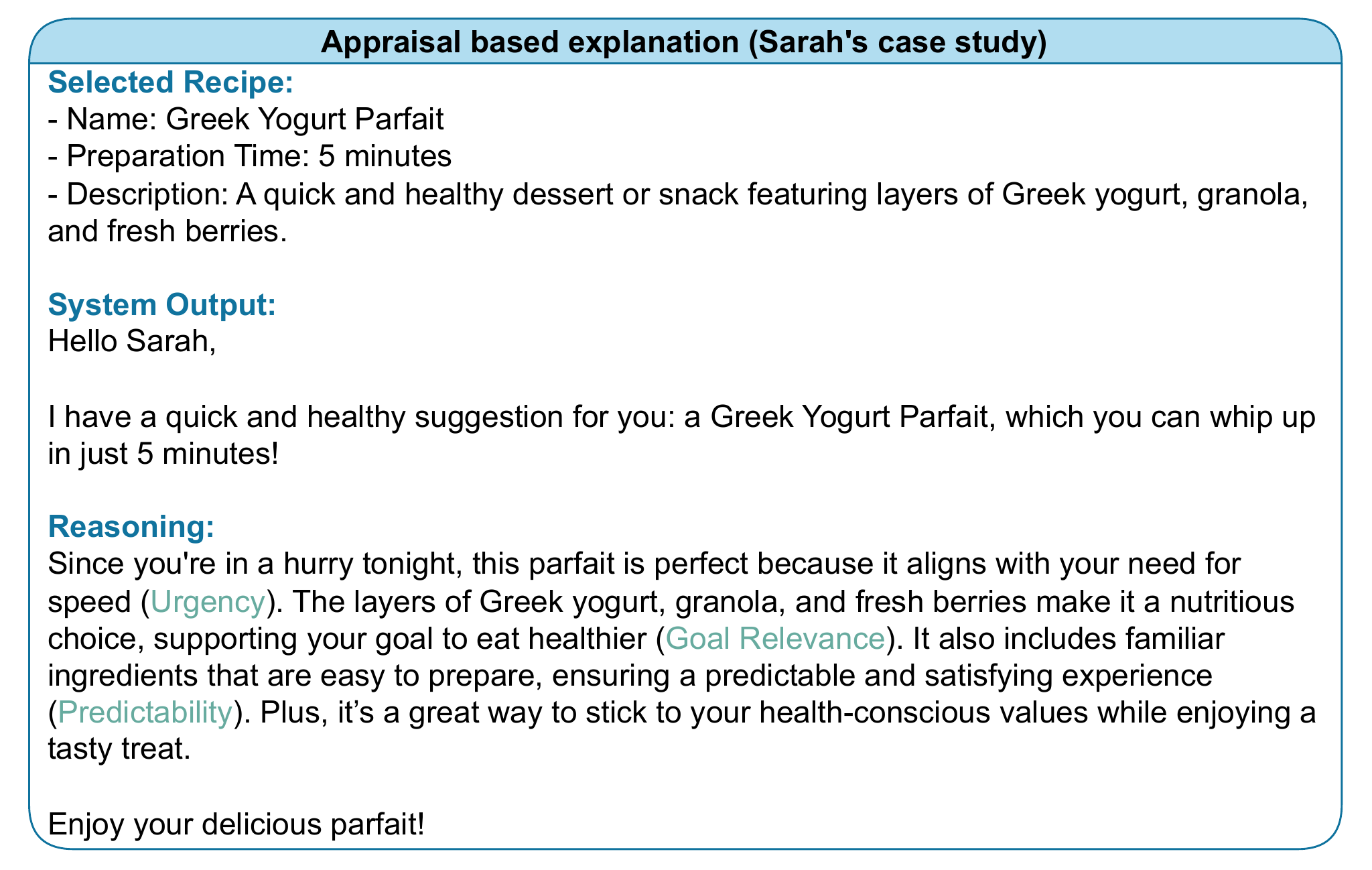}
\caption{Appraisal-based explanation for meal recommendation.}
\label{fig:Usecase1Explanation}
\end{figure}

An explanation for the top-ranked recipe is then generated using the GPT-4o model, which is prompted with the complete scenario, including user profile, situational input, dominant appraisals, and recipe details. The model is instructed to return a concise explanation that transparently justifies the recommendation, explicitly linking it to the most salient appraisal factors. As shown in Figure \ref{fig:Usecase1Explanation}, the resulting explanation is designed to be emotionally supportive, minimising cognitive load and rooted in affective reasoning. It enhances the user’s trust in the system’s behaviour by prioritising Sarah's key concerns: high urgency, goal relevance, and predictability.

\begin{figure}[htbp]
\centering
\includegraphics[width=1\linewidth]{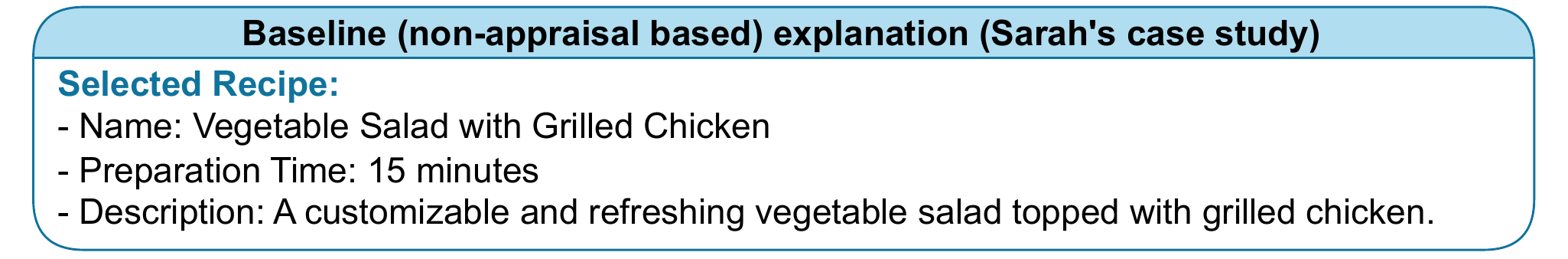}
\caption{Baseline (non-appraisal based) explanation for meal recommendation.}
\label{fig:Usecase1ExplanationControl}
\end{figure}

As shown in Figure~\ref{fig:Usecase1ExplanationControl}, we establish a baseline for this scenario to compare our approach with explanations generated using the same GPT-4o model, but without appraisal-based conditioning. In this baseline version, the model selects and explains a meal based solely on the surface-level prompt, without incorporating internal appraisal reasoning. This comparison highlights the enhanced interpretability and emotional alignment achieved by integrating CPM-based appraisals into the explanation process.

A second case study is provided in the supplementary Figures 6-8 to illustrate how the framework adapts to a different user profile and set of expectations. While these case studies focus on demonstrating the conceptual utility of appraisal for explainability, we note that the underlying NLP techniques used here are intentionally kept simple to foreground the explanatory logic of the appraisal model. Further refinement of the AI components is left for future work.

\section{Discussion}

This work presented a novel framework to bring human-centred XAI that integrates the CPM appraisal component to generate context-aware, cognitively grounded explanations in AI-driven systems. This approach contrasts with conventional XAI models, which often prioritise technical justifications, such as feature importance or model transparency, over the human-like reasoning necessary to enhance the interpretability and trustworthiness of system decisions. Focusing on appraisals such as relevance, implications, coping potential, and normative significance, our framework provides an approach that aligns with human cognitive processes, allowing for more intuitive, emotionally resonant, and user-friendly explanations.

Conventional XAI methods, such as LIME~\cite{ribeiro2016should}, SHAP~\cite{lundberg2017unified}, and counterfactual reasoning, focus heavily on technical components such as model weights or feature importance. Also, frameworks such as PeCoX’s cognitive XAI~\cite{neerincx2018using} and Belief-Desire-Intention (BDI) models~\cite{harbers2010design} adopt a human-centred approach but are limited in scope, focusing primarily on a limited set of cognitive perspectives. These models provide valuable insights into making decisions, but they often fail to bridge the gap between the system's internal processes and the human users' need for contextually relevant, meaningful explanations. As a result, they may inadvertently alienate non-expert users, limiting their ability to understand and trust system decisions. This lack of alignment with human cognition has long been a limitation in XAI, as users often seek explanations that resonate with their own emotional and cognitive evaluations, which traditional methods do not capture.

In contrast, based on the CPM’s appraisal component, our approach emphasises the cognitive dimensions that shape how humans evaluate and respond to events. Our framework provides explanations naturally aligned with human reasoning by considering appraisal dimensions such as predictability, relevance, urgency, and normative significance. For example, users can better understand a decision by knowing its rationale, how it aligns with their goals, how it might impact them emotionally, or whether it adheres to societal norms. This appraisal-based approach enhances transparency by offering explanations grounded in human thought processes and emotional responses.

The choice to focus exclusively on the CPM appraisal component has significant implications for developing user-centred XAI systems. This enhances user-centred XAI systems by aligning explanations with human judgment and decision-making. This approach evaluates system actions through cognitive dimensions, offering more profound insights into how AI decisions align with the user’s goals, emotions, and expectations. Unlike traditional XAI models that often provide post-hoc explanations, our framework delivers real-time, context-aware explanations, adapting to the user’s emotional state and situational needs. This dynamic adaptability is crucial in affective computing and human-robot interaction, where trust and transparency are key to successful collaboration.

The meal recommendation case study demonstrates our approach's practicality and effectiveness. In this scenario, the system dynamically assesses the user’s needs, priorities, and emotional state, producing context-aware explanations that resonate with the user’s current situation. Integrating appraisal dimensions such as urgency, goal relevance, and predictability allows the system to generate practically suitable and emotionally supportive recommendations. For example, Sarah’s query, `\textit{I’m hungry and in a hurry—can you make something in 15 minutes?}' triggers an explanation that addresses her immediate concerns, providing a tailored justification for the meal suggestion based on her preferences, urgency, and health-conscious goals. By comparing this appraisal-based explanation with a baseline generated without appraisal conditioning, we highlight the added value of our approach. The appraisal-conditioned explanation is more personalised, directly addressing Sarah’s emotional and cognitive context. It enhances trust in the system by prioritising the most important factors to her, making the system's behaviour more transparent and understandable. This comparison underscores the importance of considering emotional and cognitive factors in generating explanations that resonate with users, ultimately fostering stronger user-system relationships.

The framework developed in this paper has wide-ranging implications across domains like healthcare, autonomous vehicles, customer service, and personal assistants, where trust and understanding of AI decisions are essential. The system provides more accurate and emotionally attuned explanations by integrating appraisal-based reasoning, fostering greater user satisfaction, trust, and decision-making. This approach also addresses challenges in real-world XAI applications, such as misalignment with user goals, lack of contextual adaptation, and failure to consider emotional and cognitive states, making AI systems more accessible and reliable, particularly in high-stakes environments.

This work demonstrates the potential of the appraisal-based approach for improving explainability in AI systems, but several limitations remain. Scaling this framework to more complex systems could be computationally demanding, and the subjectivity of appraisal-based explanations requires validation across diverse user groups, cultural contexts, and domains. Additionally, the reliance on pre-defined appraisal dimensions may not capture all influencing factors in every scenario, so future work should explore dynamic adaptation based on user feedback and contextual factors. Integrating advanced natural language generation techniques could also enhance the fluency and naturalness of explanations. Moreover, user studies with actual participants are necessary to assess better the framework's effectiveness, which is considered future research work. These studies will provide valuable insights into how well the explanations work in practice, uncover areas for improvement, and help refine the system for real-world deployment.

Further personalisation is another key area for improvement. By incorporating user history, preferences, and cultural context, AI systems could provide more relevant and meaningful explanations, refining how they adapt to individual users. Context-specific, data-driven approaches, such as interaction data or cultural models, could ensure explanations evolve with the user's changing context, improving engagement and trust.

Finally, the simplified scoring of appraisal dimensions is a limitation of the current framework. Advanced analysers and pre-trained models, such as deep learning-based sentiment analysis, could enhance the accuracy and adaptability of emotional assessments. User studies with real participants are also necessary to evaluate the framework's effectiveness and identify areas for improvement. Despite these challenges, the approach shows promise across diverse domains, such as mental health, customer service, and education, where personalised, context-aware explanations can guide users to appropriate actions.

Overall, our synthetic use cases demonstrate the flexibility of our approach across domains like mental health monitoring, customer service, and education. By evaluating emotional states and offering tailored explanations, the system can guide users to appropriate actions, whether suggesting coping strategies for anxiety or adjusting customer interactions based on emotional feedback.

\section{Conclusion}
This paper introduces an XAI framework grounded in the appraisal component of the CPM, offering a human-centred, cognitively grounded approach to explanation generation. Our approach shows the potential to generate real-time, context-aware explanations that align with human reasoning processes by focusing on appraisal dimensions such as relevance, predictability, and normative significance. The meal recommendation case study illustrates this framework's effectiveness, providing a concrete example of how appraisal-based reasoning can improve AI systems' transparency, trust, and usability. While challenges remain, this framework presents a novel direction for developing more interpretable and emotionally resonant AI-driven systems, paving the way for better user-system interactions across various domains.

\bibliographystyle{abbrv}
\bibliography{Paper}

\begin{thebibliography}{10}

\bibitem{arnold2021explaining}
T.~Arnold, D.~Kasenberg, and M.~Scheutz.
\newblock Explaining in time: Meeting interactive standards of explanation for robotic systems.
\newblock {\em ACM Transactions on Human-Robot Interaction (THRI)}, 10(3):1--23, 2021.

\bibitem{bekler2024assessing}
M.~Bekler, M.~Yilmaz, and H.~E. Ilg{\i}n.
\newblock Assessing feature importance in eye-tracking data within virtual reality using explainable artificial intelligence techniques.
\newblock {\em Applied Sciences}, 14(14):6042, 2024.

\bibitem{cortinas2023toward}
K.~Corti{\~n}as-Lorenzo and G.~Lacey.
\newblock Toward explainable affective computing: A review.
\newblock {\em IEEE Transactions on Neural Networks and Learning Systems}, 2023.

\bibitem{delplanque2009sequential}
S.~Delplanque, D.~Grandjean, C.~Chrea, G.~Coppin, L.~Aymard, I.~Cayeux, C.~Margot, M.~I. Velazco, D.~Sander, and K.~R. Scherer.
\newblock Sequential unfolding of novelty and pleasantness appraisals of odors: evidence from facial electromyography and autonomic reactions.
\newblock {\em Emotion}, 9(3):316, 2009.

\bibitem{driggs2016communicating}
K.~Driggs-Campbell and R.~Bajcsy.
\newblock Communicating intent on the road through human-inspired control schemes.
\newblock In {\em 2016 IEEE/RSJ International Conference on Intelligent Robots and Systems (IROS)}, pages 3042--3047. IEEE, 2016.

\bibitem{giaccardi2020technology}
E.~Giaccardi and J.~Redstr{\"o}m.
\newblock Technology and more-than-human design.
\newblock {\em Design Issues}, 36(4):33--44, 2020.

\bibitem{guerdan2021toward}
L.~Guerdan, A.~Raymond, and H.~Gunes.
\newblock Toward affective xai: facial affect analysis for understanding explainable human-ai interactions.
\newblock In {\em Proceedings of the IEEE/CVF International Conference on Computer Vision}, pages 3796--3805, 2021.

\bibitem{ha2022examining}
T.~Ha, Y.~J. Sah, Y.~Park, and S.~Lee.
\newblock Examining the effects of power status of an explainable artificial intelligence system on users’ perceptions.
\newblock {\em Behaviour \& Information Technology}, 41(5):946--958, 2022.

\bibitem{harbers2010design}
M.~Harbers, K.~Van Den~Bosch, and J.-J. Meyer.
\newblock Design and evaluation of explainable bdi agents.
\newblock In {\em 2010 IEEE/WIC/ACM international conference on web intelligence and intelligent agent technology}, volume~2, pages 125--132. IEEE, 2010.

\bibitem{hilton2007course}
D.~J. Hilton and L.~M. John.
\newblock The course of events: counterfactuals, causal sequences, and explanation.
\newblock In {\em The psychology of counterfactual thinking}, pages 56--72. Routledge, 2007.

\bibitem{johnson2024explainable}
D.~S. Johnson, O.~Hakobyan, J.~Paletschek, and H.~Drimalla.
\newblock Explainable ai for audio and visual affective computing: A scoping review.
\newblock {\em IEEE Transactions on Affective Computing}, 2024.

\bibitem{kaptein2017role}
F.~Kaptein, J.~Broekens, K.~Hindriks, and M.~Neerincx.
\newblock The role of emotion in self-explanations by cognitive agents.
\newblock In {\em 2017 Seventh International Conference on Affective Computing and Intelligent Interaction Workshops and Demos (ACIIW)}, pages 88--93. IEEE, 2017.

\bibitem{kotriwala2021xai}
A.~Kotriwala, B.~Kl{\"o}pper, M.~Dix, G.~Gopalakrishnan, D.~Ziobro, and A.~Potschka.
\newblock Xai for operations in the process industry-applications, theses, and research directions.
\newblock In {\em AAAI spring symposium: combining machine learning with knowledge engineering}, pages 1--12, 2021.

\bibitem{kumar2024exploratory}
S.~Kumar, Y.~Parmet, and Y.~Edan.
\newblock Exploratory user study on verbalization of explanations.
\newblock In {\em 2024 IEEE 4th International Conference on Human-Machine Systems (ICHMS)}, pages 1--7. IEEE, 2024.

\bibitem{liu2024explainable}
Y.~Liu.
\newblock Explainable ai in eye tracking, 2024.

\bibitem{lundberg2017unified}
S.~M. Lundberg and S.-I. Lee.
\newblock A unified approach to interpreting model predictions.
\newblock {\em Advances in neural information processing systems}, 30, 2017.

\bibitem{RN590Menetrey}
M.~Q. Menétrey, G.~Mohammadi, J.~Leitão, and P.~Vuilleumier.
\newblock Emotion recognition in a multi-componential framework: The role of physiology.
\newblock {\em Frontiers in computer science}, 4:6, 2022.

\bibitem{miller2019explanation}
T.~Miller.
\newblock Explanation in artificial intelligence: Insights from the social sciences.
\newblock {\em Artificial intelligence}, 267:1--38, 2019.

\bibitem{mouakher2024explainable}
A.~Mouakher and R.~Kononov.
\newblock Explainable evaluation framework for facial expression recognition in web-based learning environments.
\newblock {\em International Journal of Machine Learning and Cybernetics}, pages 1--33, 2024.

\bibitem{neerincx2018using}
M.~A. Neerincx, J.~van~der Waa, F.~Kaptein, and J.~van Diggelen.
\newblock Using perceptual and cognitive explanations for enhanced human-agent team performance.
\newblock In {\em Engineering Psychology and Cognitive Ergonomics: 15th International Conference, EPCE 2018, Held as Part of HCI International 2018, Las Vegas, NV, USA, July 15-20, 2018, Proceedings 15}, pages 204--214. Springer, 2018.

\bibitem{ribeiro2016should}
M.~T. Ribeiro, S.~Singh, and C.~Guestrin.
\newblock " why should i trust you?" explaining the predictions of any classifier.
\newblock In {\em Proceedings of the 22nd ACM SIGKDD international conference on knowledge discovery and data mining}, pages 1135--1144, 2016.

\bibitem{sado2023explainable}
F.~Sado, C.~K. Loo, W.~S. Liew, M.~Kerzel, and S.~Wermter.
\newblock Explainable goal-driven agents and robots-a comprehensive review.
\newblock {\em ACM Computing Surveys}, 55(10):1--41, 2023.

\bibitem{sagar2024trustworthy}
S.~Sagar, A.~Taparia, H.~Mankodiya, P.~Bidare, Y.~Zhou, and R.~Senanayake.
\newblock Trustworthy conceptual explanations for neural networks in robot decision-making.
\newblock {\em arXiv preprint arXiv:2409.10733}, 2024.

\bibitem{schadenberg2021see}
B.~R. Schadenberg, D.~Reidsma, D.~K. Heylen, and V.~Evers.
\newblock “i see what you did there” understanding people’s social perception of a robot and its predictability.
\newblock {\em ACM Transactions on Human-Robot Interaction (THRI)}, 10(3):1--28, 2021.

\bibitem{RN269Scherer}
K.~Scherer, A.~Dieckmann, M.~Unfried, H.~Ellgring, and M.~Mortillaro.
\newblock Investigating appraisal-driven facial expression and inference in emotion communication.
\newblock {\em Emotion}, 2019.

\bibitem{RN21Scherer}
K.~R. Scherer.
\newblock The dynamic architecture of emotion: Evidence for the component process model.
\newblock {\em Cognition and Emotion}, 23(7):1307--1351, 2009.

\bibitem{scherer2013coregrid}
K.~R. Scherer, J.~R.~F. Fontaine, and C.~Soriano.
\newblock {\em CoreGRID and MiniGRID: Development and validation of two short versions of the GRID instrument1}.
\newblock Oxford University Press, Oxford, 2013.

\bibitem{SomarathnaPerCom9767281}
R.~Somarathna, T.~Bednarz, and G.~Mohammadi.
\newblock An exploratory analysis of interactive vr-based framework for multi-componential analysis of emotion.
\newblock In {\em 2022 IEEE International Conference on Pervasive Computing and Communications Workshops and other Affiliated Events (PerCom Workshops)}, pages 353--358, 2022.

\bibitem{somarathnaexploring}
R.~Somarathna and G.~Mohammadi.
\newblock Exploring the component process model of emotions using interactive vr games.
\newblock {\em Intelligent Computing}.

\bibitem{somarathna2022multiAJCAI}
R.~Somarathna, A.~Quigley, and G.~Mohammadi.
\newblock Multi-componential emotion recognition in vr using physiological signals.
\newblock In {\em Australasian Joint Conference on Artificial Intelligence}, pages 599--613. Springer, 2022.

\bibitem{somarathna2023emostim}
R.~Somarathna, P.~Vuilleumier, and G.~Mohammadi.
\newblock Emostim: A database of emotional film clips with discrete and componential assessment.
\newblock {\em IEEE Transactions on Affective Computing}, pages 1--11, 2023.

\bibitem{suffian2023investigating}
M.~Suffian, I.~Stepin, J.~M. Alonso-Moral, A.~Bogliolo, et~al.
\newblock Investigating human-centered perspectives in explainable artificial intelligence.
\newblock In {\em CEUR Workshop Proceedings}, volume 3518, pages 47--66, 2023.

\bibitem{RN34Reekum}
C.~van Reekum, T.~Johnstone, R.~Banse, A.~Etter, T.~Wehrle, and K.~Scherer.
\newblock Psychophysiological responses to appraisal dimensions in a computer game.
\newblock {\em Cognition and Emotion}, 18(5):663--688, 2004.

\bibitem{zhan2023evaluating}
H.~Zhan, D.~C. Ong, and J.~J. Li.
\newblock Evaluating subjective cognitive appraisals of emotions from large language models.
\newblock {\em arXiv preprint arXiv:2310.14389}, 2023.

\end{thebibliography}

\clearpage
\onecolumn
\begin{center}
\section*{Supplementary Materials}

\begin{figure*}[ht]
\centering
\includegraphics[width=0.85\linewidth]{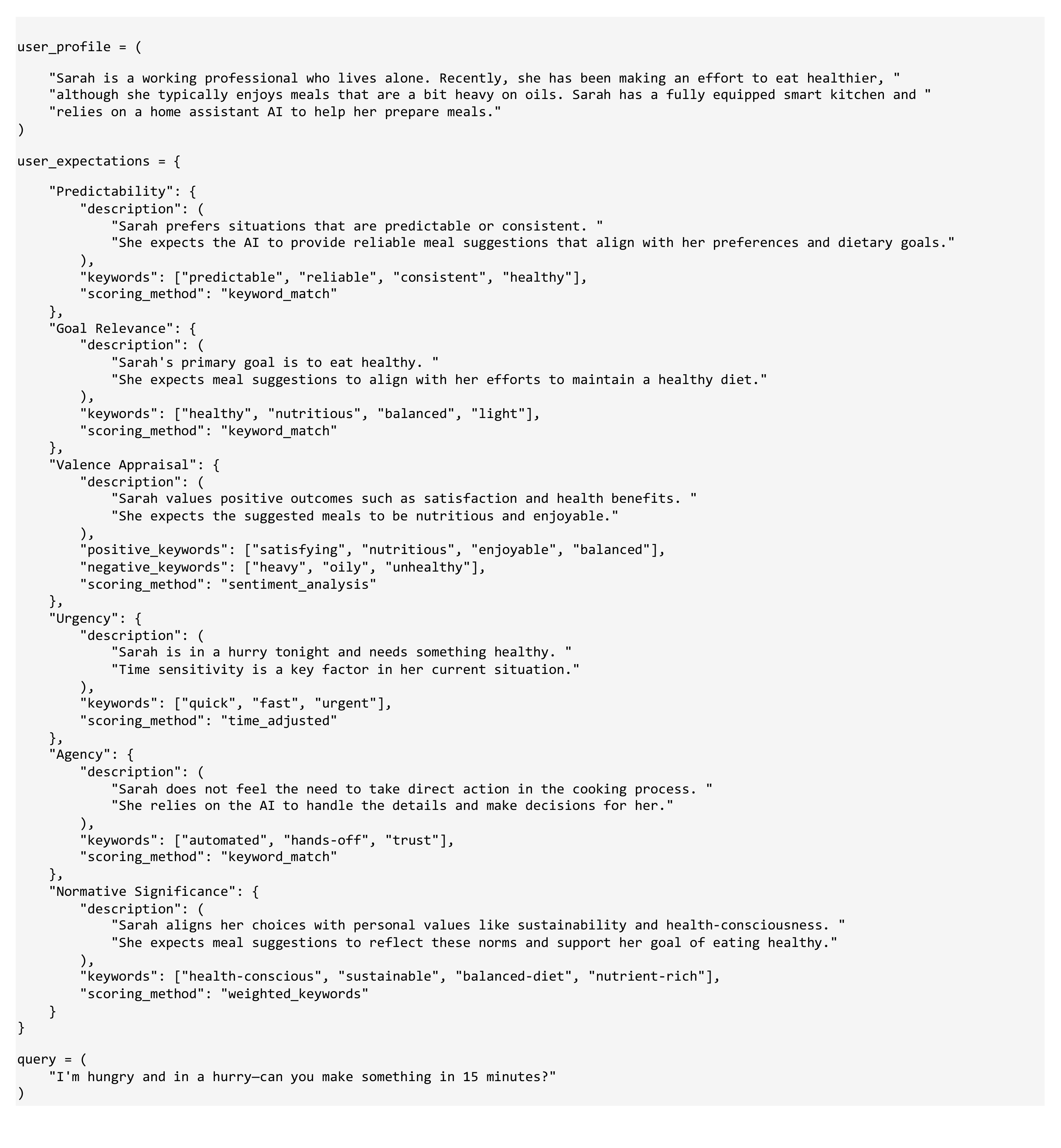}
\caption{The input data and prompts used in Sarah's case study are illustrated in the diagram, highlighting how various appraisal dimensions—such as predictability, goal relevance, and urgency—impact meal suggestions tailored to Sarah's preferences. The keywords used for scoring, such as \textit{healthy, quick}, and \textit{nutritious}, were simulated to align with the selected scoring methods, as simulated data were used for this case study.}
\label{fig:UseCase1Prompts}
\end{figure*}

\begin{figure*}
\includegraphics[width=1\linewidth]{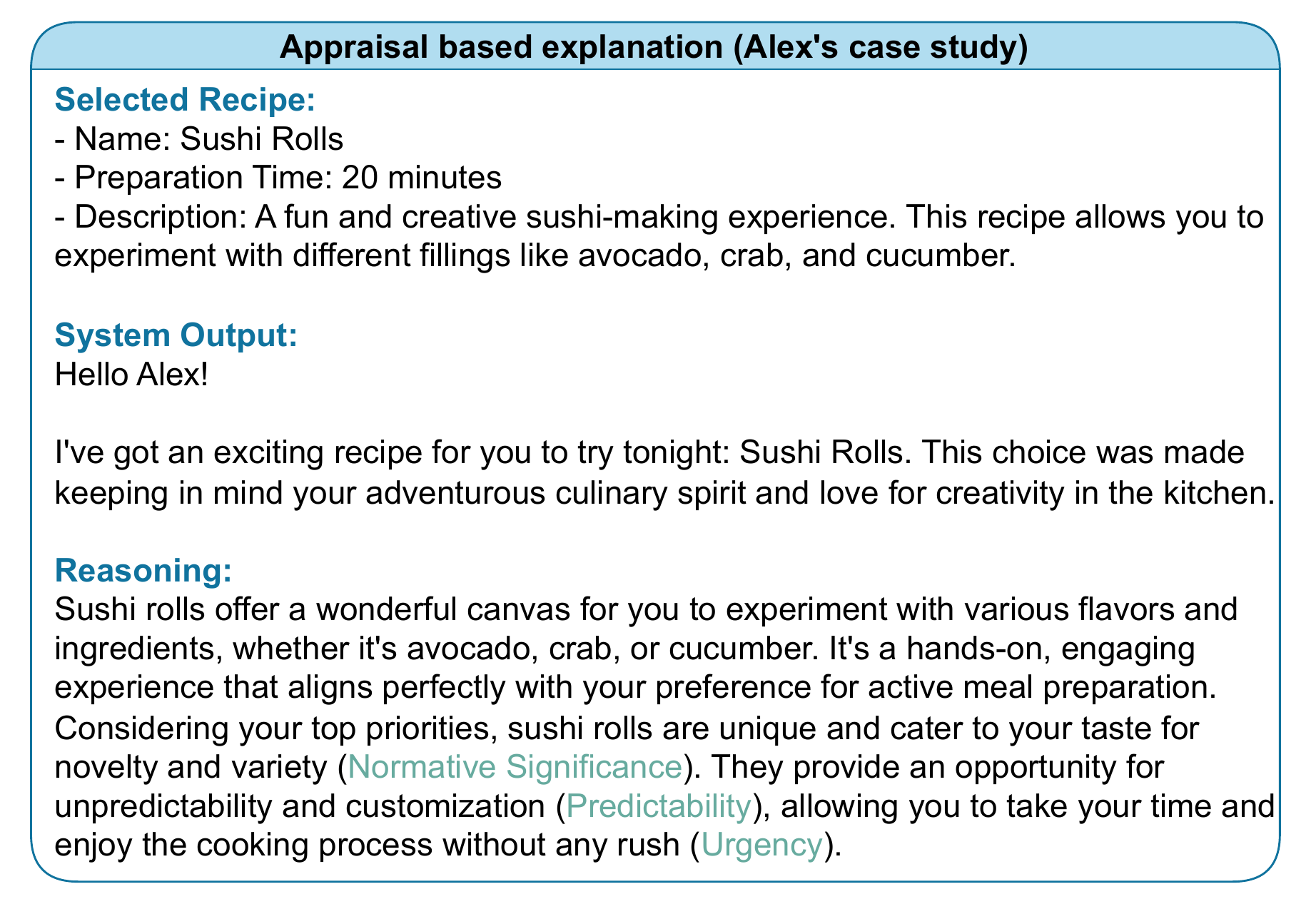}
\caption{Appraisal-based explanation for meal recommendation.}
\label{fig:Usecase2Explanation}
\end{figure*}

\begin{figure*}
\centering
\includegraphics[width=0.8\linewidth]{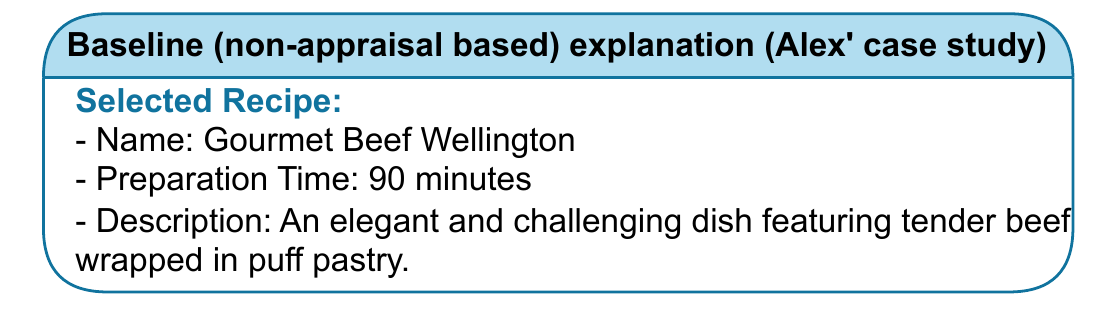}
\caption{Baseline (non-appraisal based) explanation for meal recommendation.}
\label{fig:Usecase2ExplanationControl}
\end{figure*}

\begin{figure*}[htbp]
\centering
\includegraphics[width=1\linewidth]{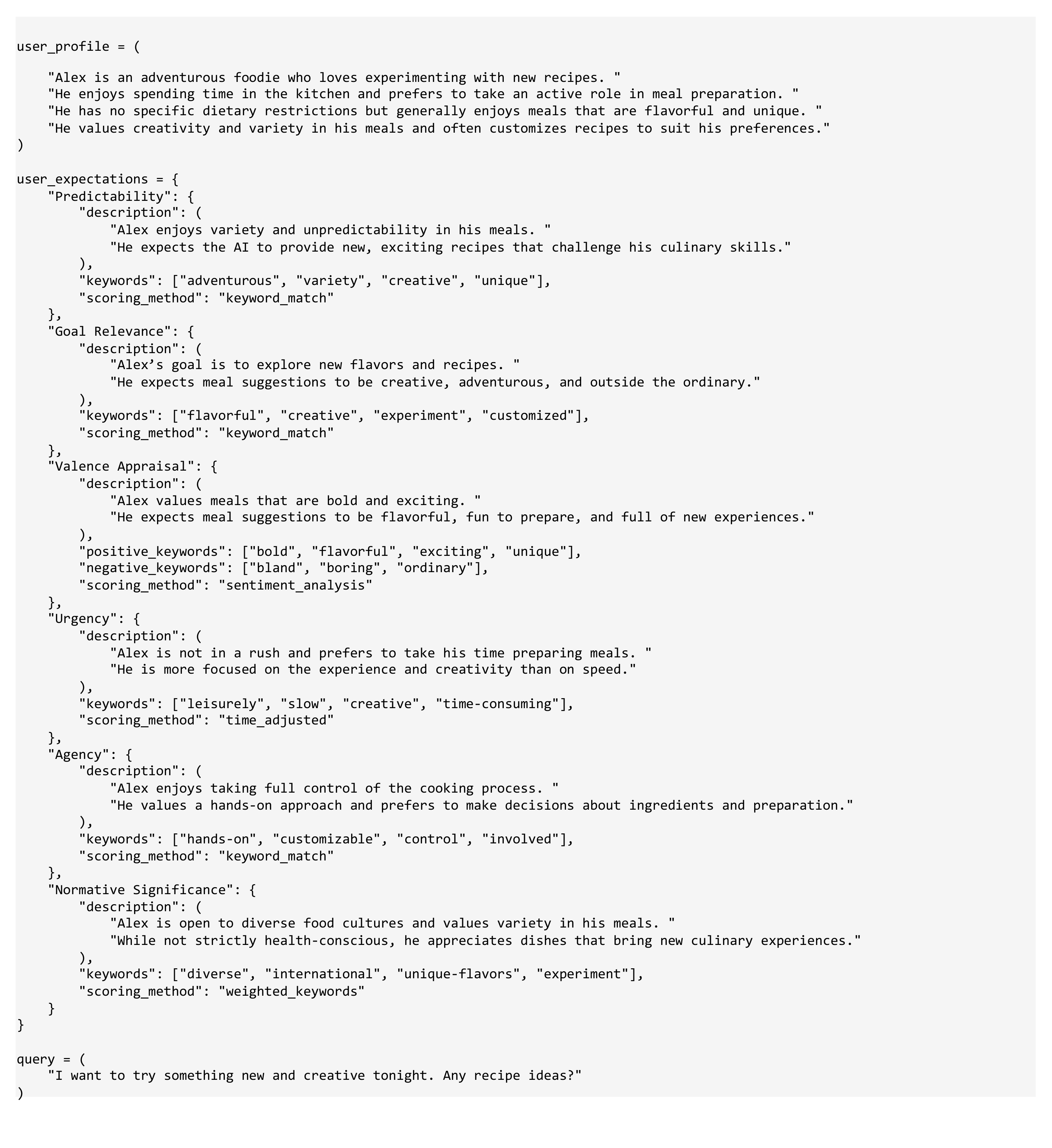}
\caption{The input data and prompts used in Alex's case study are illustrated in the diagram, highlighting how various appraisal dimensions—such as predictability, goal relevance, and urgency—impact meal suggestions tailored to Alex's preferences. The keywords used for scoring, such as \textit{adventurous, customized}, and \textit{exciting}, were simulated to align with the selected scoring methods, as simulated data were used for this case study.}
\label{fig:UseCase2Prompts}
\end{figure*}
\end{center}

\end{document}